\documentclass[twocolumn,preprintnumbers,msmath,amssymb,prl,floatfix,final]{revtex4-2}
\usepackage{graphicx}
\usepackage{dcolumn}
\usepackage{float}
\usepackage{bm}
\usepackage{siunitx}
\usepackage{units}
\usepackage{amsmath}
\usepackage{mciteplus}
\usepackage[utf8]{inputenc}
\makeatletter 
\def\@citess#1{\textsuperscript{[#1]}}
\makeatother 
\usepackage{comment}

\begin{document}

\newcommand{\ittext}[1]{\mbox{\rm\scriptsize #1}}

\title{Tip-Induced $\beta$-Hydrogen Dissociation in an Alkyl Group Bound on Si(001)}

\author{A.~Adamkiewicz$^{1}$, T.~Bohamud$^{1}$, M.~Reutzel$^{1,2}$, U.~H{\"o}fer$^{1}$, and M.~D\"urr$^{1,3,*}$}
\address{$^1$Fachbereich Physik and Zentrum f{\"u}r Materialwissenschaften, Philipps-Universit{\"a}t Marburg, D-35032 Marburg, Germany\\
$^2$I.~Physikalisches Institut, Georg-August-Universit\"at G\"ottingen, D-37077 G\"ottingen, Germany\\
$^3$Institut f{\"u}r Angewandte Physik and Zentrum f\"ur Materialforschung,
Justus-Liebig-Universit{\"a}t Giessen, D-35392 Giessen, Germany\\
$^*$corresponding author: Michael D\"urr, E-mail:
michael.duerr@ap.physik.uni-giessen.de}

\hyphenation{tempera-ture}

\begin{abstract}

{\sl {\bf Abstract:} Atomic-scale chemical modification of surface-adsorbed ethyl groups on Si(001) was induced and studied by means of scanning tunneling microscopy. Tunneling at sample bias $>$~+1.5\,V leads to tip-induced C-H cleavage of a $\beta$-hydrogen of the covalently bound ethyl configuration. The reaction is characterized by the formation of an additional Si-H and a Si-C bond. The reaction probability shows a linear dependence on the tunneling current at 300\,K; the reaction is largely suppressed at 50\,K. The observed tip-induced surface reaction at room temperature is thus attributed to a one-electron excitation in combination with thermal activation.\\}

\end{abstract}

\maketitle

\section{Introduction}

Implementing surface chemistry on the nanoscale in the design and fabrication process of miniaturized electronic devices may open a wide range of possible applications~\cite{Flood04sci,Aradhya13natnano,Sun14csr}. By investigating both the properties of the adsorption process and the nature of surface reactions of adsorbed molecules, controlled functionalization of semiconductor surfaces, in particular of the technologically most important Si(001) surface, can be realized~\cite{Yates98sci,Wolkow99arpc,Hamers00acr,Filler03pss,Yoshin04pss,Leftwich08ssr,Schofield13natcomm,Reutzel16jpcc,Laenger19jpcm,Glaser20jpcc}.\\ \indent
One method to manipulate adsorbates on the molecular level is scanning tunneling microscopy (STM)~\cite{Kim15pss,Dujardin16ss,MacLean17jacs,Okabayashi18pnas,Kazuma18sci,Rusimova18sci,Mette19acie}. Often, these studies focus on tip-induced lateral or vertical movement of an adsorbed species~\cite{Kuhnle02ss,Komeda02sci,Ueda03srlett,Grill08jpcm,Bohamud20jpcc} or the reaction from an intermediate into a covalently bound final state~\cite{Mo92prl,Hla00prl,Mette19acie}.

Here, we study a complex tip-induced reaction scheme, starting from a covalently bound adsorbate on Si(001) (Fig. 1): Ethyl species on Si(001), which are thermally stable at room temperature, were manipulated by tip-induced excitation resulting in new covalently bound final states. They are characterized by a Si-C-C-Si entity accompanied by a Si-H bond involving a Si atom of a neighbouring dimer. Tip-induced reaction takes place for an applied sample bias of $U_{\text{t}}$~$>$~1.5\,V; above the threshold voltage, the reaction probability depends linearly on the tunneling current, indicating a one-electron excitation mechanism. At 50~K, the reaction is largely suppressed. Thus, the dominant process at 300\,K is assigned to a combination of electronic and thermal excitation.

\begin{figure}[b]
\begin{center}
\includegraphics[width=\columnwidth]{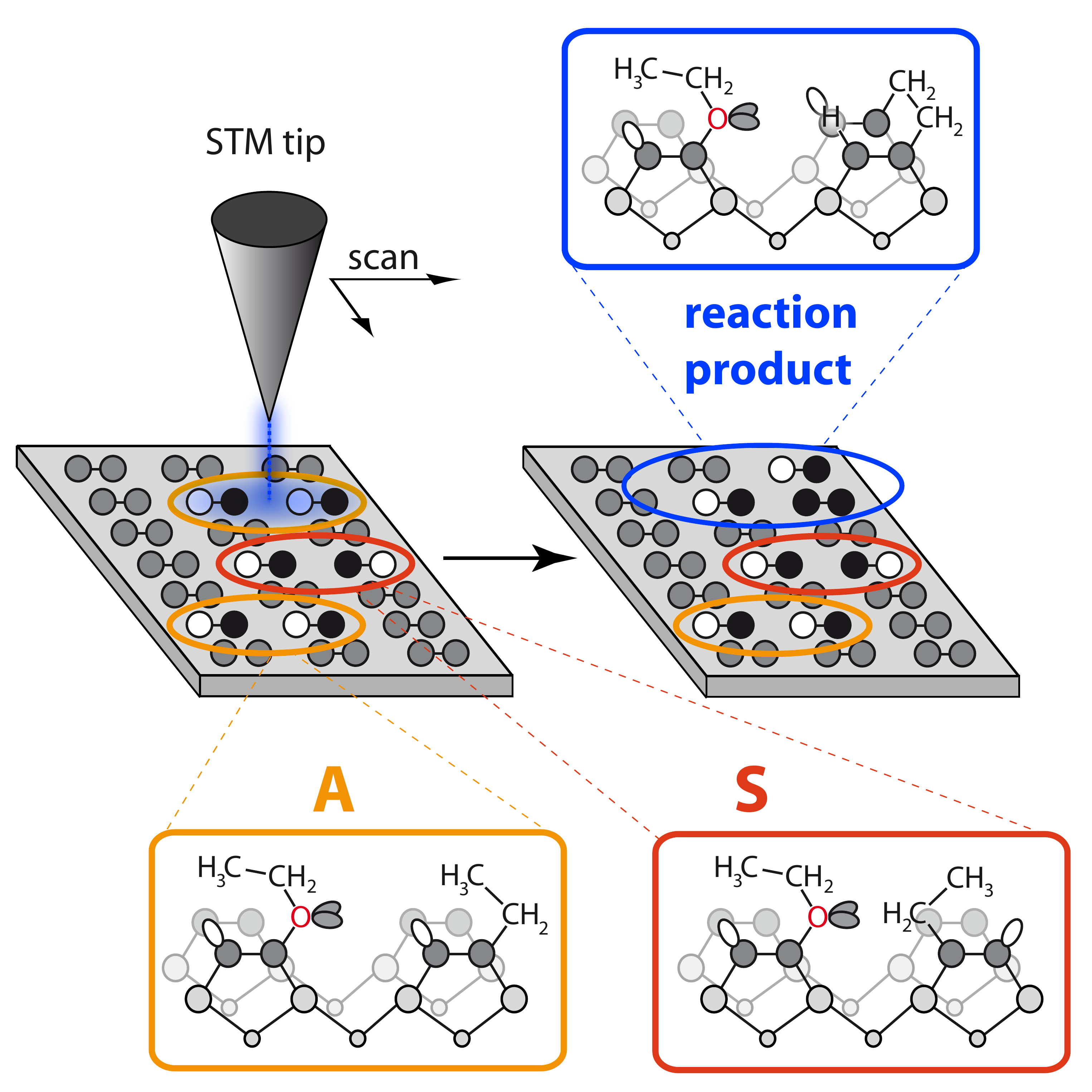}
\caption{Schematic representation of the two covalently bound adsorption configurations of diethyl ether on Si(001) after ether cleavage, asymmetric (A) and symmetric (S), and the tip-induced reaction product. The A and S configurations represent the starting configurations of this study. Tip-induced excitation leads to a surface reaction involving the ethyl species of these configurations. The resulting reaction products extend over three Si dimers.}
\end{center}
\end{figure}

\section{Experimental Section}

The experiments were performed in an ultrahigh vacuum chamber (OMICRON VT-STM) with a base pressure $<$\,1$\times$10$^{\text{-10}}$\,mbar. The \textit{n}-doped Si samples were oriented within 0.25$^{\circ}$ along the (001) direction. Pristine sample surfaces were obtained by degassing at 700\,K and applying repeated heating cycles to surface temperatures above 1450\,K. With cooling rates of about 1\,K$/$s, a well ordered 2$\times$1 reconstruction was obtained~\cite{Schwalb07prb,Mette09cpl}. The low temperature measurements were performed by cooling the sample with liquid helium to a surface temperature of $T_{\text{s}}$~$\approx$~50\,K.\\ \indent
Diethyl ether was dosed via a leak valve up to a coverage of 0.01\,ML (1\,ML represents one diethyl ether molecule per Si dimer). The adsorption of the ether molecules proceeds via a datively bound intermediate state~\cite{Reutzel15jpcc,Reutzel15jpcl}. At room temperature, thermal activation induces ether cleavage, resulting in two possible final states, symmetric (S) and asymmetric (A) (Fig.~1). These configurations are characterized by -OCH$_{\text{2}}$CH$_{\text{3}}$ and -CH$_{\text{2}}$CH$_{\text{3}}$ fragments bound to Si dimers of neighbouring dimer rows, either next to each other (configuration S) or with one Si atom in between (configuration A)~\cite{Reutzel15jpcc}. The ethyl entity of these configurations serves as the starting point of the investigated surface reaction.

\begin{figure}[b!]
\begin{center}
\includegraphics[width=\columnwidth]{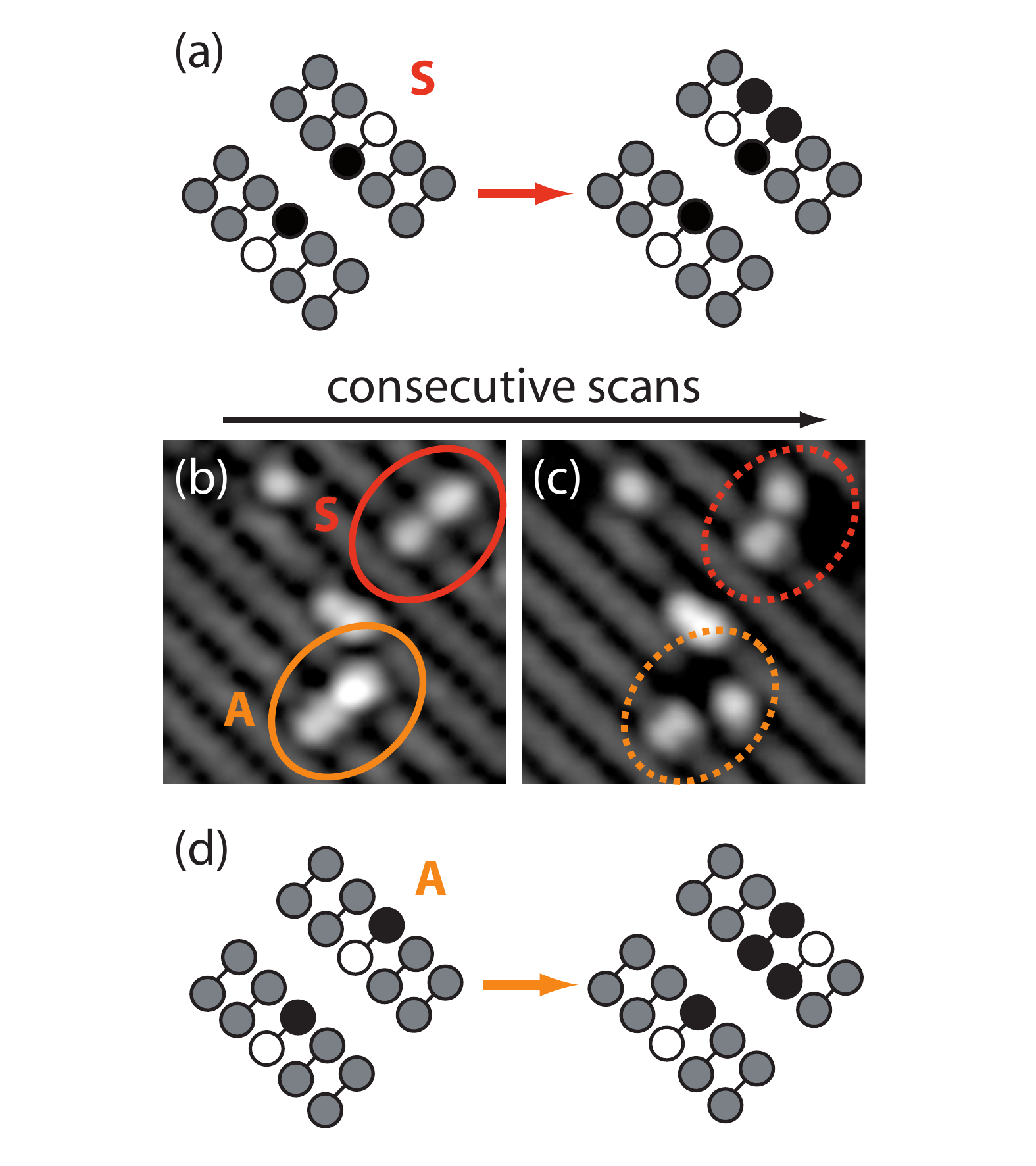}
\caption[fig:2]{STM images (6.4 $\times$ 6.4\,nm$^{\text{2}}$, $U_{\text{t}}$~=~0.8\,V, $I_{\text{t}}$~=~2.0\,nA, $T_{\text{s}}$~$=$~300\,K) of diethyl ether reacted on Si(001). (b) shows the area before and (c) after a scan at 2.5\,V and 2.0\,nA. Before the reaction, each configuration comprises two dimers (b). The reaction products extend over three Si dimers (dashed ellipses in (c)). In the 2D schemes in (a) and (d), the configurations before and after the 2.5-V-scan are shown. Black and white circles represent saturated Si atoms and isolated dangling bonds, respectively.}
\label{fig:2}
\end{center}
\vspace{-5mm}
\end{figure}

\begin{figure}[b!]
\begin{center}
\includegraphics[width=\columnwidth]{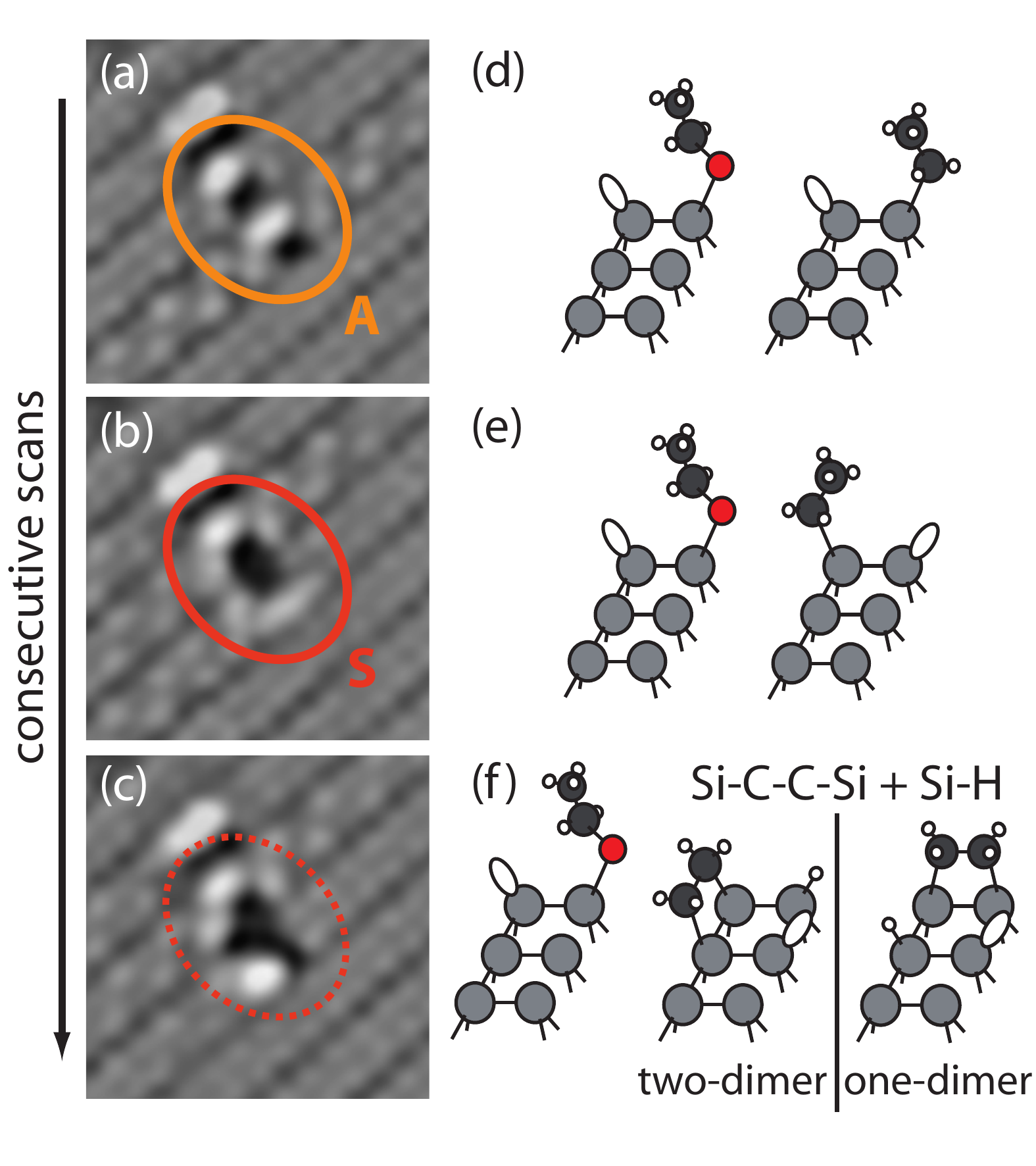}
\caption[fig:3]{Consecutive STM images (4.5 $\times$ 4.5\,nm$^{\text{2}}$, $U_{\text{t}}$~=~2.3\,V, $I_{\text{t}}$~=~0.5\,nA, $T_{\text{s}}$~$=$~300\,K) showing the conversion from asymmetric configuration A in (a) to symmetric configuration S in (b), followed by a tip-induced surface reaction leading to the final configuration in (c). In the schemes (d) to (f), red, dark grey, and small white circles represent O, C, and H atoms, respectively. Isolated dangling bonds are depicted as white ellipses. The final product is shown with the C$_{\text{2}}$H$_{\text{4}}$ unit in a two-dimer (left) or one-dimer configuration (right). For details see main text).}
\label{fig:3}
\end{center}
\vspace{-5mm}
\end{figure}

\section{Results}

In Fig.~2, a pair of subsequent STM images is shown, the change of the observed configurations indicates tip-induced reactions of the adsorbates on the surface. In detail, Fig.~2(b) shows the empty state STM image of Si(001) after diethyl ether adsorption at room temperature which is characterized by two bright features on neighbouring dimer rows; a symmetric and an asymmetric configuration which originate from thermally activated ether cleavage can be identified. The bright features of the configurations observed in the STM image correspond to isolated dangling bonds on a dimer of which the second silicon atom is reacted by one of the molecular fragments~\cite{Durr13pss,Reutzel15jpcc}. Scanning the area with $U_{\text{t}}$~=~2.5\,V and $I_{\text{t}}$~=~2.0\,nA led to tip-induced surface reactions as indicated in (c). In both cases, one of the diethyl ether fragments is manipulated, resulting in a feature extending over two dimers of the same dimer row: a suppression of dangling bonds at the dimer where the brighter feature was originally located and a bright feature on a neighbouring dimer atom of the same dimer row. The dimension along the row of this tip-induced feature is 4.2$\pm$0.5\,$\mathring{\text{A}}$, in good comparison to the distance between two Si dimers of 3.85\,$\mathring{\text{A}}$~\cite{Hamers86prb}.

STM images of the reaction product that explicitly show the suppression of two additional dangling bonds are shown in Fig.~3 and Fig.~S1 of the Supporting Information. Since in the final products two more Si atoms are saturated when compared to the initial configurations S and A, a mere movement of the molecular fragment can be excluded. The dark area at the position of the dimer on which the ethyl fragment was bound prior to excitation is interpreted as two saturated Si atoms. The single bright feature on the neighbouring dimer can be understood as an isolated dangling bond of one dimer atom which arises due to the saturation of the second dimer atom in the course of the tip-induced reaction~\cite{Mette19acie,Durr13pss,Reutzel15jpcc}.

In Fig.~2(b), the isolated dangling bond signatures of the asymmetric configuration are of unequal brightness. For A-configurations, the tip-induced reaction was only observed for the fragment corresponding to the brighter dangling bond feature, which was assigned to the ethyl entity based on its position on the dimer and taking into account the reaction pathway of diethyl ether on Si(001). A detailed discussion of this assignment is found in Ref.~\cite{Reutzel15jpcc}. The sequence of STM images shown in Fig.~3 also allows identification of the different parts of the symmetric configuration shown in Fig.~3(b) as it starts from an asymmetric configuration as shown in Fig.~3(a). From (a) to (b), the configuration has changed due to hopping of the ethyl fragment on-top of one dimer, leading to the observed symmetric configuration~\cite{Bohamud20jpcc}. In Fig.~3(c), the configuration is observed after a surface reaction has been induced. Thus, also in this case, starting from a symmetric configuration, we can clearly assign the reaction to the Si-CH$_{\text{2}}$CH$_{\text{3}}$ entity. Taking all observations into account, we conclude that the observed tip-induced surface reaction is exclusively featuring the ethyl fragment of both the S and A configurations. This conclusion holds for both the experiments at 300\,K and 50\,K.

As two further dangling bonds are saturated in the final configuration, a covalent bond in the adsorbed species has to be broken. Taking into account the geometric constraints, the most obvious scenario involves tip-induced cleavage of the C-H bond of a $\beta$-hydrogen atom, which results in an additional Si-H and Si-C bond. Two configurations are then compatible with the number and spatial distribution of quenched dangling bonds as sketched in Fig.~3(f): Either with the C$_{\text{2}}$H$_{\text{4}}$ unit on-top of one dimer or in the so-called endbridge configuration on two neighbored dimers. Both configurations are known to be stable adsorption configurations of C$_{\text{2}}$H$_{\text{4}}$ (ethylene) on Si(001) \cite{Mette09cpl,Akagi16ss,Pecher17cpc}. In line with the on-top configuration is a further configuration which consists of four saturated dangling bonds in a row (Fig.~S2). This configuration, which was rarely observed, can be interpreted as a one-dimer Si-CH$_2$-CH$_2$-Si species with the additional hydrogen atom being bound to the closest silicon atom of the {\sl next} dimer row.

The mechanism of the observed tip-induced surface reaction was further investigated by means of the dependence of the reaction probability on tip-sample bias and tunneling current. In Fig.~4, the bias dependent measurements of the reaction probability are shown for $T_{\rm s} = 300$~K. The data reveal a threshold voltage of $U_{\text{th}}$~$\approx$~1.5\,V. Above the threshold, the reaction probability shows a nonlinear dependence on bias voltage. The dependence for $I_{\text{t}}$~=~0.5\,nA and 2.0\,nA is comparable when the data are normalized with respect to the applied values of the tunneling current (Fig.~4, inset).

\begin{figure}[t!]
\vspace{-1mm}
\begin{center}
\includegraphics[width=0.95\columnwidth]{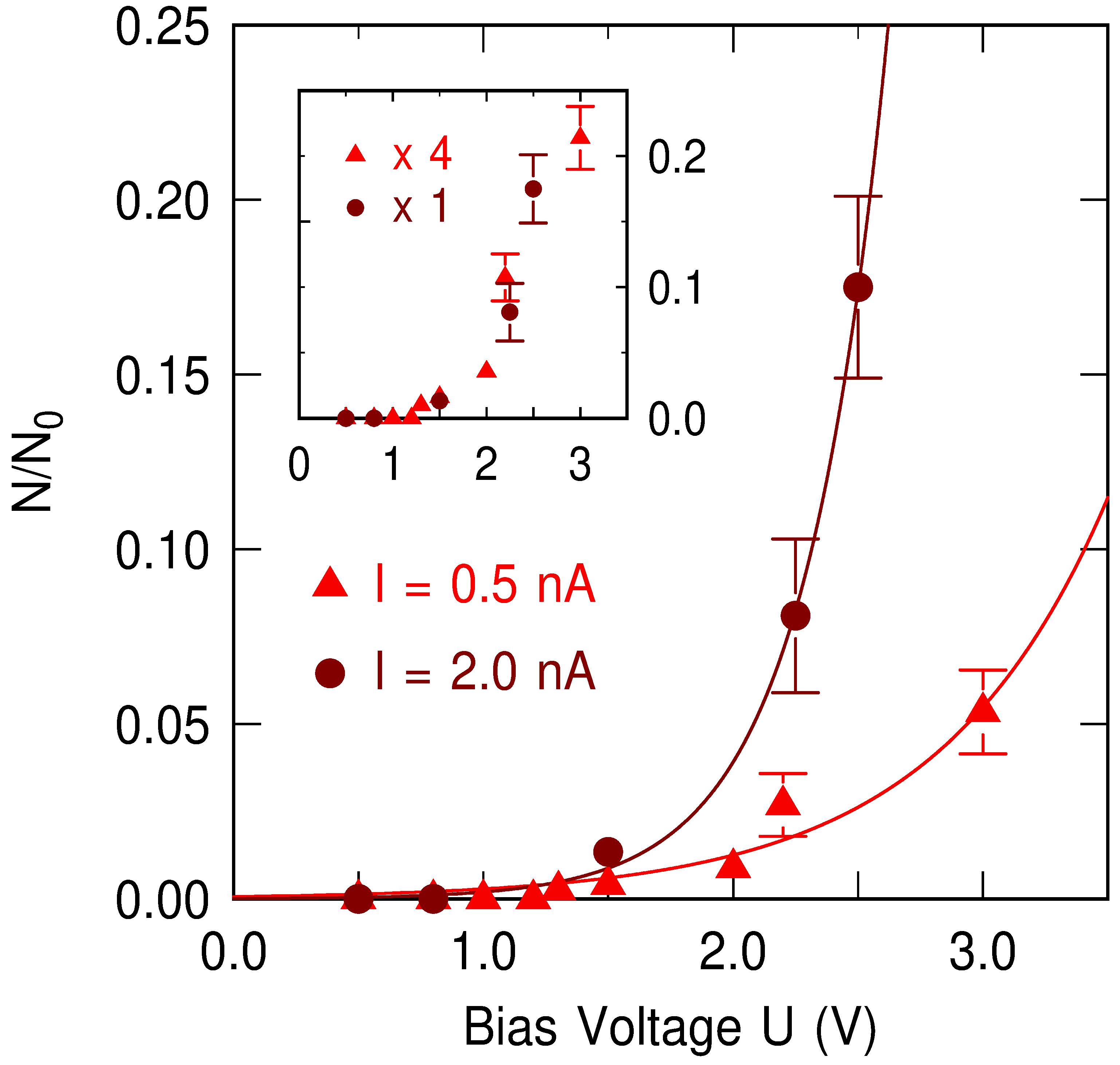}
\caption[fig:4]{Reaction probability in terms of the relative number of conversions $N / N_0$ ($N$: number of conversions, $N_0$: initial number of configurations) after one scan as a function of positive bias voltage at $T_{\text{s}}$~$=$~300\,K for $I_{\text{t}}$~=~0.5\,nA and $I_{\text{t}}$ = 2.0\,nA. Symbols: data; lines: exponential fits. The higher tunneling current results in an overall increased probability for the conversion process. In the inset, the data for $I_{\text{t}}$~=~0.5\,nA are multiplied by a factor of 4 to account for the reduced number of tunneling electrons in comparison to the 2.0-nA-data.}
\end{center}
\vspace{4mm}
\end{figure}

In Fig.~5, the reaction probability is shown as a function of the tunneling current for $U_{\text{t}} = 2.5$~V. At 300\,K, a linear dependence on the tunneling current is observed, indicating a one-electron process. With an estimated radius of 0.5~nm for the area in which the electronic excitation can induce the observed reaction, we calculate an efficiency of about $10^{-10}$ reactions per tunneling electron. At 50\,K surface temperature, a significantly lower and constant reaction probability was observed in the same range of tunneling current values (Fig.~5).

\begin{figure}[t!]
\begin{center}
\includegraphics[width=0.95\columnwidth]{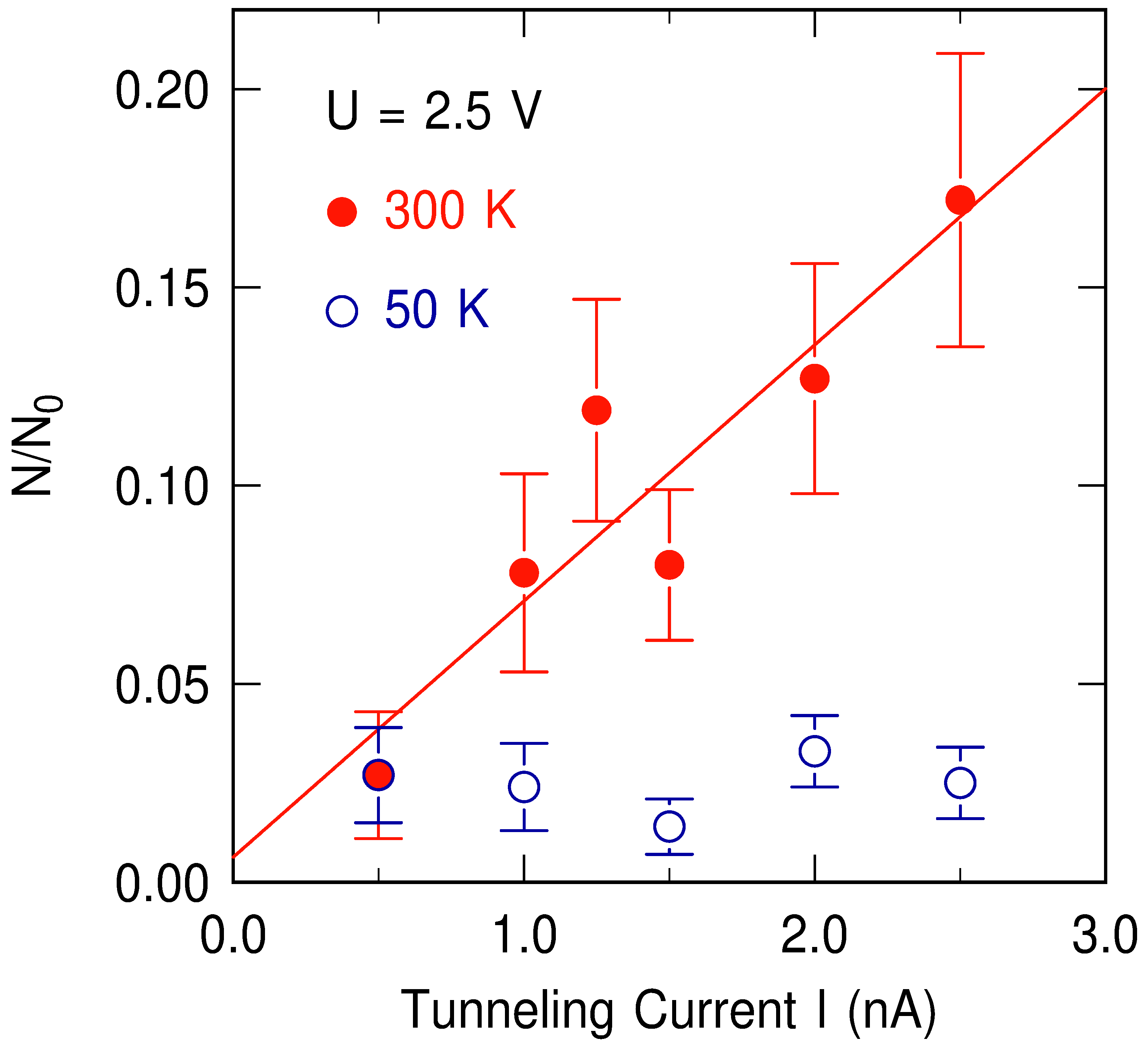}
\caption[fig:5]{Relative number of conversions $N/N_0$ after one scan as a function of the tunneling current at 2.5\,V sample bias for $T_{\text{s}}$~$=$~300\,K and 50\,K. The red line shows a linear fit to the data at 300\,K.}
\label{fig:5}
\end{center}
\vspace{0mm}
\end{figure}

\section{Discussion}

In the following, we want to discuss and further interpret the experimental findings on the tip-induced reaction of a surface-adsorbed ethyl entity on Si(001). First, the observed threshold voltage of $U_{\text{th}}$~$\approx$~1.5\,V in the data of Fig.~4 is too small for direct electron injection into molecular states of the adsorbates taking into account a large HOMO-LUMO separation and level alignment similar to related systems \cite{Mette14cpc,Mette19acie,Mao19sc}. Therefore, the tip-induced process is likely to involve electrons tunneling into surface states of the substrate.

However, this excitation is only effective in the close vicinity of the adsorbed ethyl group: when we analyze the raster scans at increased bias which initiated the tip-induced reaction, many of the so-called split-features can be observed (Fig.~S3 in the Supporting Information): These split-features are signatures which are divided into two parts; they show one part of the configuration in the state before the reaction took place and the second part of the configuration in the state after the reaction has occurred. Thus they indicate that the reaction took place with the tip being localized close to the configuration. The ratio of split-features to non-split-features at 300\,K is 0.7:1, pointing towards a small radius of the area around the adsorbed ethyl groups in which tip-induced excitation of the surface states is effective in inducing the observed reaction.

The interpretation of a substrate-mediated one-electron excitation mechanism via electron injection into substrate surface states is further supported by the fact that no reaction processes were observed with negative sample bias at 300\,K.

The largely suppressed reactivity at 50\,K indicates that, in addition to the electronic excitation, thermal activation is necessary for this tip-induced surface reaction. Thermally assisted one-electron excitation leading to bond dissociation was also reported in the context of other systems, e.g., in the case of C-Cl bond cleavage of chlorobenzene on Si(111)~\cite{Sakulsermsuk10acsnano}.

The remaining, low reaction probability observed at 50\,K does not show any dependence on the number of incident electrons. It is attributed to a minority species of adsorbates, for which the reaction can be induced with much less thermal activation.
According to the number of configurations reacted at 50~K, this minority species accounts for less than 3~\% of the initial configurations. Field-induced effects, which are also largely independent of the tunneling current \cite{Bohamud20jpcc}, can be ruled out: also at 50~K, we observe the process to take place only with the tip very close to the adsorbate, in contrast to the non-local character of field-induced processes \cite{Bohamud20jpcc,Alemani06jacs,Calupitan17jpcc}.

The tip-induced reaction studied here for ethyl groups on Si(001), which are part of the reaction products of diethyl ether on Si(001), was not observed in the case of tetrahydrofuran reacted on Si(001)~\cite{Mette19acie}. This might be a result of the more restricted configuration of the $\beta$-carbon atom due to the dual-tethered final configuration of tetrahydrofuran after ether cleavage on Si(001), which bridges two dimer rows \cite{Mette14cpc}. In contrast, the -CH$_{\text{2}}$CH$_{\text{3}}$ fragment studied here has a high degree of freedom in terms of molecular configurations. This gives further indication that the H and C atoms have to be in a certain configuration, most likely in close vicinity to the surface, in order to efficiently interact with the dangling bonds as a prerequisite for the tip-induced reaction to occur. 

The latter might be associated with $\beta$-hydride elimination well-known in organic chemistry, induced by electron injection into a close-by dangling bond state and followed by a [2+2] cycloaddition of the generated C=C double bond. Independent of the details of the mechanism, however, a concerted process is most likely for the observed surface reaction, comprising the dangling bonds, the $\beta$-carbon, and the reacting $\beta$-hydrogen.

\section{Conclusions}

We have shown a tip-induced C-H cleavage at the $\beta$-carbon of ethyl groups adsorbed on Si(001).
The process was found to be substrate-mediated and at 300\,K, the underlying mechanism is a combination of one-electron excitation and thermal activation. In a most likely scenario, the tunneling electrons are injected into the dangling bond of the respective Si atom while thermal activation of the molecular adsorbates and/or the neighbored dimers leads to the reactive configuration, altogether enabling C-H cleavage at surface-adsorbed C$_{\text{2}}$H$_{\text{5}}$ species. Starting from a stable, covalently bound configuration, the process allows to induce new configurations which can be controlled on the atomic scale by means of temperature, bias voltage, and tunneling current.

\section{Supplementary Material}

The Supplementary Material includes STM images of the tip-induced final configurations taken at various samples bias, STM images of a configuration which extends over three dimer rows, and an STM image illustrating a ''split feature``.

\section{Acknowledgment}

We gratefully acknowledge funding by Deutsche Forschungsgemeinschaft through DU~1157/4-1, GRK 1782, and SFB 1083 (project-ID 223848855).

\bibliographystyle{prsty}

\bibliography{abkz,agof,hsiutf8,tpputf8,new_library}

\end{document}


\renewcommand{\thepage}{S\arabic{page}}
	\renewcommand{\thefigure}{S\arabic{figure}}
	\renewcommand{\thetable}{S\arabic{table}}

	\setstretch{1.6}
	\noindent Supporting Information for:
	\begin{center}
		\begin{large}
			\textbf{Tip-Induced $\beta$-Hydrogen Dissociation in an Alkyl Group Bound on Si(001)}\\
		\end{large}
		\vspace{3mm}
		\begin{small}
		A.~Adamkiewicz$^{1}$, T.~Bohamud$^{1}$, M.~Reutzel$^{1,2}$, U.~H{\"o}fer$^{1}$, and M.~D\"urr$^{1,3,*}$\\
		
$^1$\emph{Fachbereich Physik and Zentrum f{\"u}r Materialwissenschaften,\\ Philipps-Universit{\"a}t Marburg, D-35032 Marburg, Germany}\\
$^2$\emph{I.~Physikalisches Institut, Georg-August-Universit\"at G\"ottingen, D-37077 G\"ottingen, Germany}\\
$^3$\emph{Institut f{\"u}r Angewandte Physik and Zentrum f\"ur Materialforschung,\\
Justus-Liebig-Universit{\"a}t Giessen, D-35392 Giessen, Germany}\\
$^*$\emph{Corresponding author: michael.duerr@ap.physik.uni-giessen.de}
		\end{small}

		\date{\today}

	\end{center}
	\setstretch{1.3}
	\vspace{30mm}
	This Supporting Information includes:
	\begin{itemize}
		\item[\textbf{(I)}] STM images showing the tip-induced configuration of surface-adsorbed ether groups for different tip-sample bias
        \item[\textbf{(II)}] STM images showing a rarely observed tip-induced reaction product
		\item[\textbf{(III)}] STM image showing a so-called ''split feature``
        		
		\end{itemize}
	
	\pagebreak

\noindent \textbf{I.~STM images showing the tip-induced configuration of surface-adsorbed ether groups for different tip-sample bias}\\

\begin{figure}[h!]
	\begin{center}
		\includegraphics[width = 7cm]{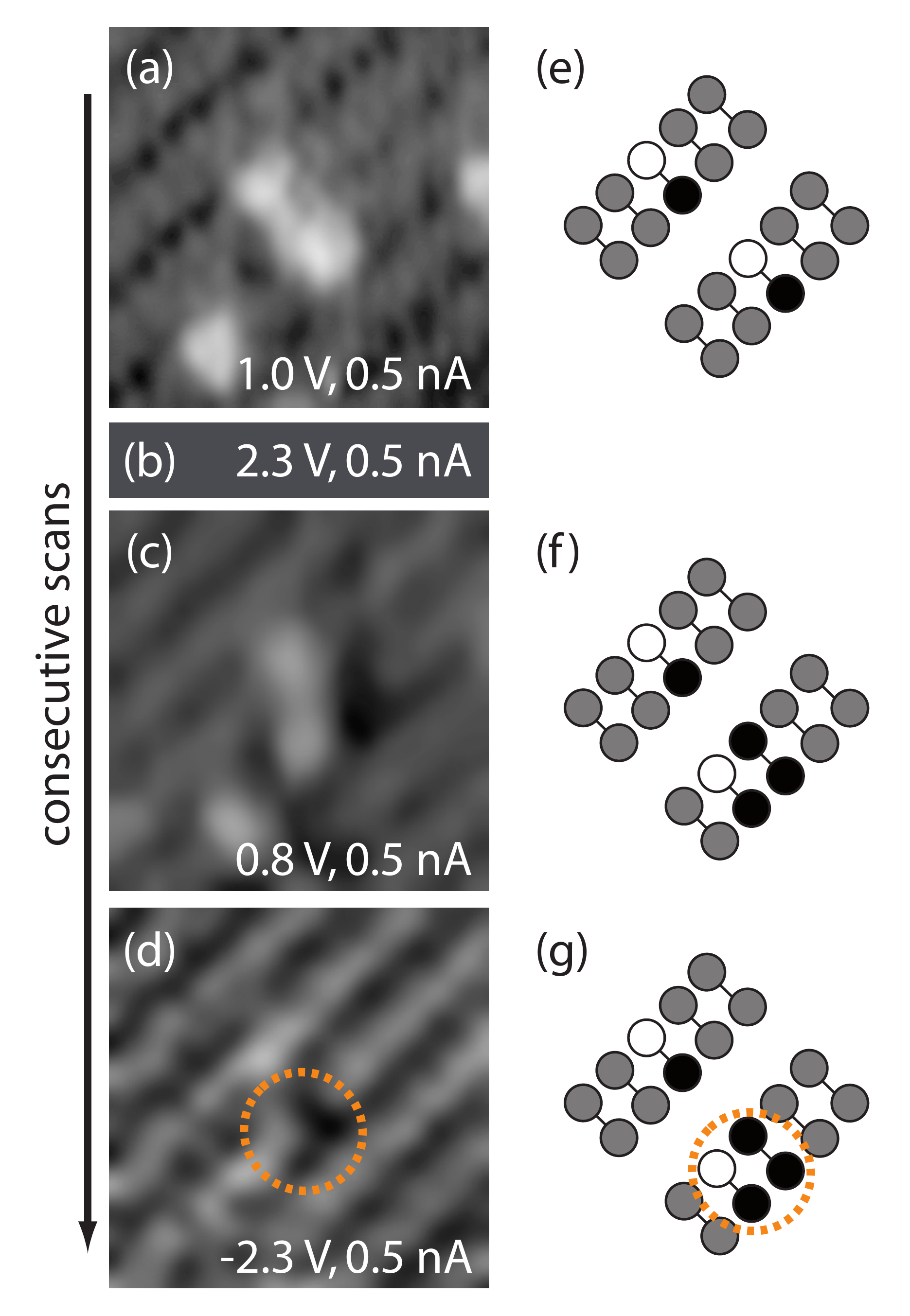}
		\caption{Consecutive STM images (4.5 $\times$ 4.5\,nm$^{\text{2}}$, $T_{\text{s}}$~$=$~300\,K) showing the initial and the final configuration of the tip-induced conversion. The final configuration is shown at positive and negative sample bias. (a) shows the asymmetric configuration of diethyl ether reacted on Si(001)~\cite{Reutzel15jpcc}. A scan with $U_{\text{t}}$~=~2.3\,V, $I_{\text{t}}$~=~0.5\,nA (b) caused a surface reaction, resulting in the configuration shown in (c) and (d) at positive and negative sample bias, respectively. The reaction product comprises two additional suppressed dangling bonds when compared to the initial configuration, which is in particular evident in (d) where the contribution of the isolated dangling bonds to the tunneling current is less pronounced than in the empty state image (c). In the 2D schemes in (e) to (g), black and white circles represent saturated Si atoms and isolated dangling bonds, respectively.
			 \label{S1}}
	\end{center}
\end{figure}

\newpage

\noindent \textbf{II.~STM images showing a rarely observed tip-induced reaction product}\\
	
	\begin{figure}[h!]
		\begin{center}
		\includegraphics[width = 11cm]{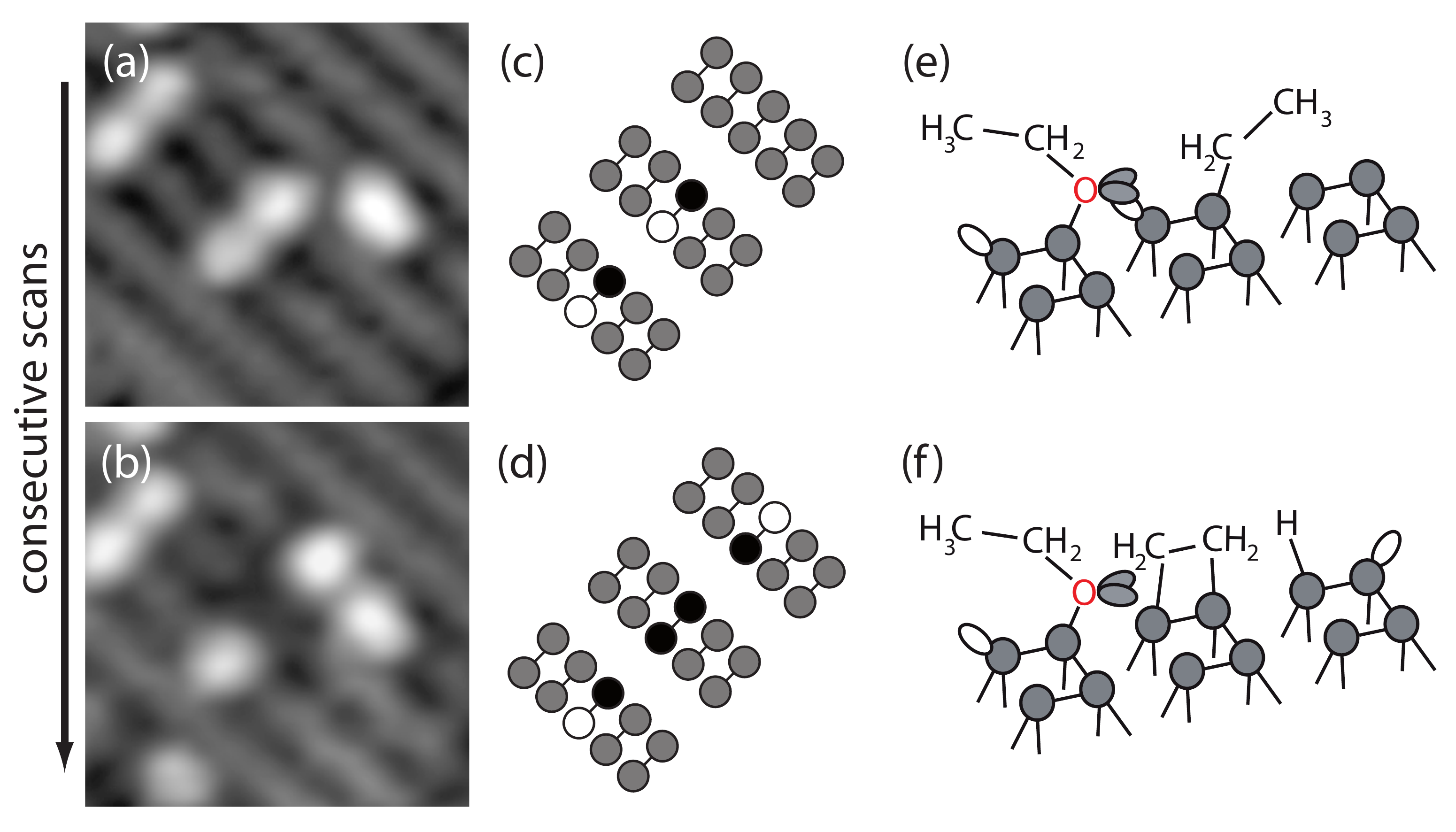}
		\caption{(a) and (b): Consecutive STM images (6 $\times$ 6\,nm$^{\text{2}}$, $U_{\text{t}}$~=~0.8\,V, $I_{\text{t}}$~=~0.5\,nA, $T_{\text{s}}$~$=$~300\,K) and corresponding 2D ((c) and (d)) and 3D schemes ((e) and (f)) showing a rarely observed tip-induced reaction product. (a) shows the asymmetric configuration of diethyl ether reacted on Si(001)~\cite{Reutzel15jpcc} and (b) the configuration after a single scan with $U_{\text{t}}$~=~2.5\,V, $I_{\text{t}}$~=~0.5\,nA. The reaction product consists of four saturated dangling bonds in a row (d). Assuming that the Si-C bond of the former Si-CH$_{\text{2}}$CH$_{\text{3}}$ entity remains intact, the configuration can be interpreted as a one-dimer Si-C-C-Si species with the hydrogen atom being bound to a dimer of the next dimer row (f). In the 2D schemes in (c) and (d), black and white circles represent saturated Si atoms and isolated dangling bonds, respectively.}
		\label{S2}
		\end{center}
	\end{figure}

\newpage

\noindent \textbf{III.~STM image showing a so-called ''split feature``}\\
	
	\begin{figure}[h!]
		\begin{center}
		\includegraphics[width = 11cm]{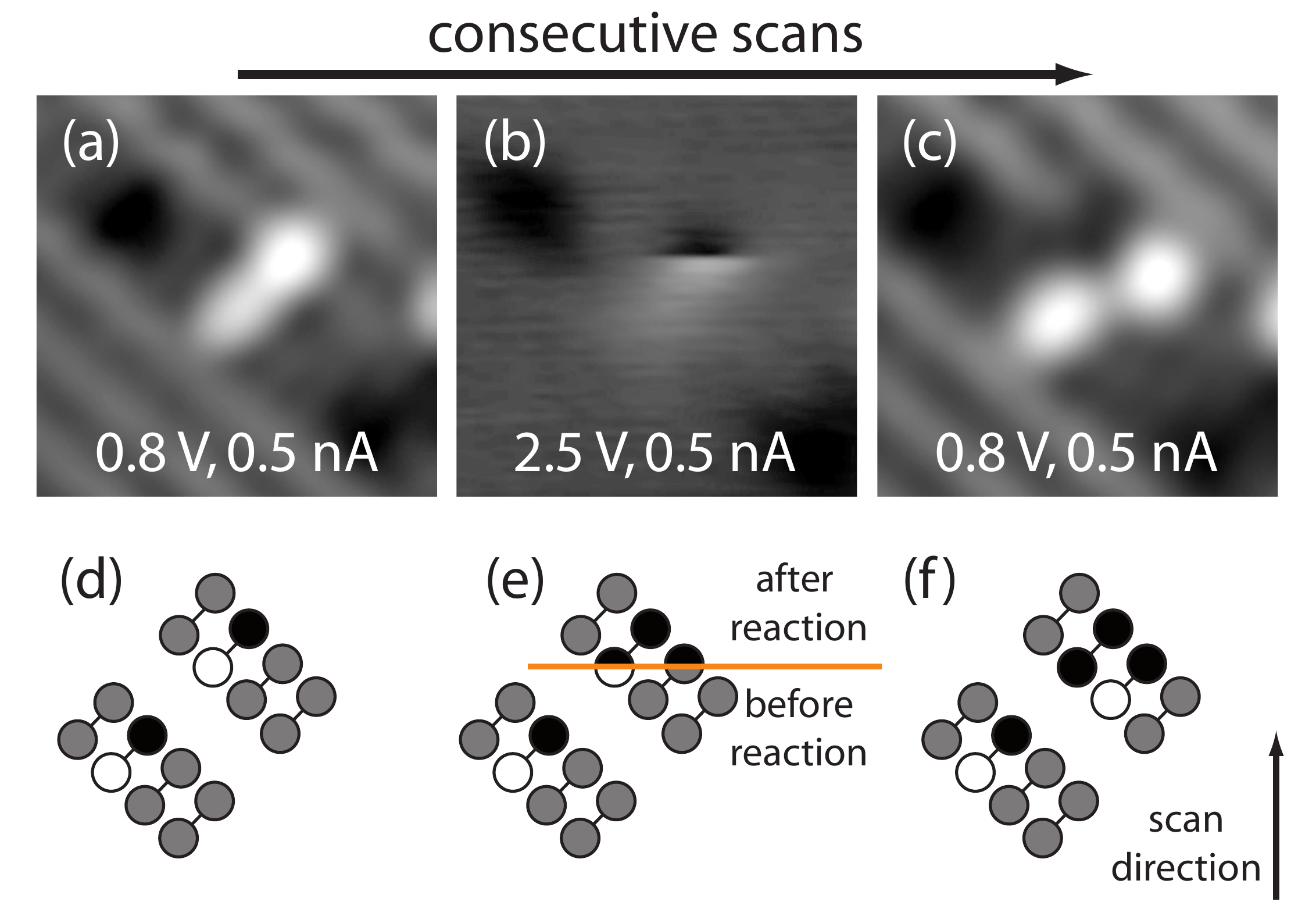}
		\caption{Consecutive STM images ((a) to (c), 4.1 $\times$ 4.1\,nm$^{\text{2}}$, $T_{\text{s}}$~$=$~300\,K) and respective 2D schemes of the surface ((d) to (f), black and white circles represent saturated Si atoms and isolated dangling bonds, respectively). In (b), a tip-induced reaction takes place when the tip is scanning a line which crosses the adsorbate (horizontal scan direction of the single lines, order of lines from bottom to top). In the lower part of (b), the image shows the adsorbate in the initial configuration, in the upper part, the adsorbate is imaged in the final configuration. The conversion has occurred when scanning the line which is indicated in orange in (e). }
		\label{S2}
		\end{center}
	\end{figure}

\bibliographystyle{prsty}

\bibliography{abkz,agof,hsiutf8,tpputf8,new_library}